\documentclass[crop,authoryear]{CSL}

\usepackage{amsmath}
\usepackage{amssymb}
\usepackage{graphicx}
\usepackage{booktabs}
\usepackage{array}
\usepackage{tikz}
\usetikzlibrary{positioning}

\newcommand{\fitwidth}[1]{\resizebox{\ifdim\width>\columnwidth\columnwidth\else\width\fi}{!}{#1}}
\newcolumntype{P}[1]{>{\raggedright\arraybackslash}p{#1}}

\jname{International Journal of Astrobiology}

\begin{document}

\supertitle{Research Paper}

\title[Prospects for Panspermia via ISOs]{Prospects for Panspermia via Interstellar Objects like 3I/ATLAS}

\author[Kakharov and Loeb]{Shokhruz Kakharov$^{1}$ and Abraham Loeb$^{1}$}

\address{\add{1}{Astronomy Department, Harvard University, 60 Garden St., Cambridge, MA 02138, USA}}

\corres{\name{Shokhruz Kakharov} \email{shokhruzbekkakharov@college.harvard.edu}}

\begin{abstract}
We study the feasibility of natural and directed panspermia via interstellar objects (ISOs) like 3I/ATLAS. The paper is organized around two questions. First, could natural panspermia occur if microbes or biomolecules survived inside shielded ice and were later exposed during perihelion and outbound activity? Second, could directed panspermia occur if a technological civilization planted life-bearing material inside or onto an icy ISO so that it later transported life through the Milky Way?

We combine data on 3I/ATLAS with order-of-magnitude thermal, biological, and mission constraints. SPHEREx provides the volatile and organic context through CO$_2$, H$_2$O, CO, dust, and a broad C--H feature, while JWST/MIRI provides the first direct CH$_4$ detection in an interstellar object and confirms an unusual volatile inventory, including enhanced CO$_2$:H$_2$O and CH$_4$:H$_2$O ratios. We distinguish dormant interstellar cruise from active perihelion.

Natural panspermia is plausible as microbes can survive or repair damage in ice films, veins, or frozen matrices at very low metabolic rates. Methane production is more nuanced. Frozen survival metabolism would require $\sim10^{14}$--$10^{15}$ kg of biomass to match the JWST CH$_4$ rates, but active methanogenic archaea in warm, liquid, substrate-rich settings can produce methane many orders of magnitude faster, reducing the required biomass in optimistic laboratory-rate comparisons. Directed panspermia faces a different challenge: a direct 60 km s$^{-1}$ impact releases $1.8\times10^9$ J kg$^{-1}$, hundreds of times the specific energy of TNT, and would destroy a biological capsule.

3I/ATLAS-like objects are therefore best treated as test cases for panspermia diagnostics rather than as evidence for life. Natural panspermia requires preservation plus a credible liquid-water or near-surface activation pathway. Directed panspermia requires gentle emplacement, shielding, and payload thermal control; present-day radiogenics cannot make 100 m--1 km icy ISOs broadly habitable.
\end{abstract}


\selfcitation{Kakharov S and Loeb A. Prospects for Panspermia via Interstellar Objects like 3I/ATLAS.}

\received{--}
\revised{--}
\accepted{--}

\maketitle

\section{Introduction}

Macroscopic interstellar objects (ISOs) are now an observed population of extrasolar bodies passing through the Solar System, with 1I/`Oumuamua and 2I/Borisov establishing the observational class and motivating future surveys \citep{jewittseligman2023}. Their discovery has also made ISO-mediated panspermia a question that can be constrained by observations, thermal physics, and mission design \citep{ginsburg2018}. 1I/`Oumuamua showed that solid bodies from other planetary systems pass through the Solar System \citep{meech2017}. 2I/Borisov showed a more familiar comet-like case \citep{guzik2020}. 3I/ATLAS is especially important because it is an active, volatile-rich ISO with observed CO$_2$, H$_2$O, CO, refractory dust, organic C--H emission, and a direct post-perihelion CH$_4$ detection by JWST/MIRI.

This paper is not primarily an ISO survey-yield study. It asks whether objects like 3I/ATLAS can participate in panspermia, including transfer mediated by interstellar objects \citep{lingamloeb2018}. We split that question into two parts. The first is \emph{natural panspermia}: could microbes or biomolecules have survived inside a shielded icy ISO for interstellar times, and could methane production during the active perihelion/outbound phase include any contribution from microbial activity? The second is \emph{directed panspermia}: could a technological civilization deliberately place a life-bearing capsule or biological payload into an icy ISO when it passes near a star, allowing the object to transport life through the Galaxy? The latter is a version of directed panspermia in the sense of deliberate biological seeding \citep{crickorgel1973}.

Detection rates, warning times, and impactor calculations are useful only insofar as they constrain panspermia prospects. The central issue is whether the thermal, chemical, and dynamical environment permits survival, revival, or deliberate emplacement of life-bearing material.

The distinction between dormant cruise and active perihelion is essential. A dormant ISO in interstellar space can have a cold surface controlled by the interstellar environment. Near perihelion, however, 3I/ATLAS is not a cold body: solar heating drives sublimation, coma formation, dust release, and exposure of deeper volatile reservoirs. Panspermia scenarios must therefore specify whether they require long-term frozen survival, short-lived near-surface liquid films, a sustained liquid-water interior, or an engineered capsule.

\Fpagebreak

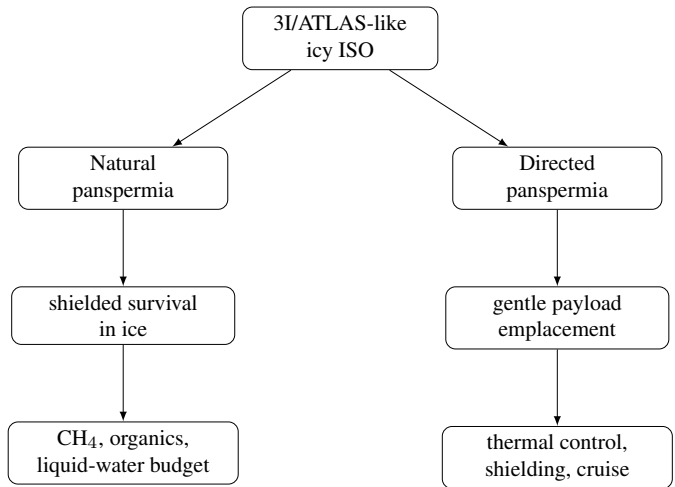
\begin{figure}[ht!]
  \centering
  \resizebox{\columnwidth}{!}{%
  \begin{tikzpicture}[>=latex, node distance=1.2cm, every node/.style={font=\small}]
    \node[draw, rounded corners, align=center, minimum width=2.8cm, minimum height=0.9cm] (iso) {3I/ATLAS-like\\icy ISO};
    \node[draw, rounded corners, align=center, below left=1.1cm and 0.2cm of iso, minimum width=3.0cm] (natural) {Natural\\panspermia};
    \node[draw, rounded corners, align=center, below right=1.1cm and 0.2cm of iso, minimum width=3.0cm] (directed) {Directed\\panspermia};
    \node[draw, rounded corners, align=center, below=1.1cm of natural, minimum width=3.2cm] (survival) {shielded survival\\in ice};
    \node[draw, rounded corners, align=center, below=1.1cm of directed, minimum width=3.2cm] (payload) {gentle payload\\emplacement};
    \node[draw, rounded corners, align=center, below=1.1cm of survival, minimum width=3.3cm] (test1) {CH$_4$, organics,\\liquid-water budget};
    \node[draw, rounded corners, align=center, below=1.1cm of payload, minimum width=3.3cm] (test2) {thermal control,\\shielding, cruise};
    \draw[->] (iso) -- (natural);
    \draw[->] (iso) -- (directed);
    \draw[->] (natural) -- (survival);
    \draw[->] (directed) -- (payload);
    \draw[->] (survival) -- (test1);
    \draw[->] (payload) -- (test2);
  \end{tikzpicture}}
  \caption{Conceptual split used in this paper. Natural panspermia asks whether an ISO can preserve cells or biomolecules and later expose them during perihelion or outbound activity. Directed panspermia asks whether a payload can be positioned gently and then use the ISO as shielding and interstellar transport.}
  \label{fig:panspermia_concept}
\end{figure}

\section{3I/ATLAS as the observational test case}

3I/ATLAS provides the observational motivation for this study. Pre-perihelion SPHEREx observations showed a CO$_2$-rich coma with water ice/water gas and limited or tentative CO. Post-perihelion SPHEREx observations showed stronger H$_2$O, CO, CO$_2$, CN, refractory dust, and a broad organic C--H feature at 3.2--3.6 $\mu$m \citep{lisse2026spherex}. The important wording is that SPHEREx detected a broad C--H organic feature, not a unique isolated methane line. JWST/MIRI later reported the first direct CH$_4$ detection on an ISO, with $Q_{\rm CH4}\approx 4.2\times10^{26}$ molecules s$^{-1}$ on 2025 December 15--16 at 2.20 AU and $2.3\times10^{26}$ molecules s$^{-1}$ on 2025 December 27 at 2.54 AU \citep{belyakov2026jwst}.

JWST/MIRI directly detected methane for the first time on an interstellar visitor, found CH$_4$ surprisingly high relative to water, confirmed that the object remained unusually CO$_2$-rich relative to water, and observed water production declining fastest as the comet moved outbound from the Sun, making 3I/ATLAS a useful panspermia test case \citep{belyakov2026jwst}. The same work gives the preferred nonbiological interpretation: delayed methane is consistent with CH$_4$ buried below the processed surface and released only after solar heating reached deeper icy layers.

The timing of the volatiles matters. CO$_2$ activity appearing early can be explained if CO$_2$-rich surface or near-surface layers, CO$_2$-bearing grains, or mixed volatile phases were already accessible before perihelion. Methane is more volatile than CO$_2$, so a delayed CH$_4$ detection may seem surprising, but it does not require biology. Possible abiotic explanations include depletion of surface methane by earlier heating or cosmic-ray processing, burial below a processed mantle, trapping in clathrates or mixed ices, diffusion barriers in porous ice, production from irradiated organics, or exposure of fresher subsurface material after perihelion.

Although CH$_4$ and CO$_2$ are biologically relevant molecules, their detection alone is not a biosignature. The relevant test is whether their abundance, timing, isotopic composition, spatial distribution, and association with organics are inconsistent with abiotic volatile release. The panspermia potential of 3I/ATLAS is therefore not that methane proves life. Rather, it supplies a concrete set of diagnostics: volatile timing, spatial morphology, isotopic ratios, organic composition, chirality, and whether organics are associated with dust grains, icy grains, or gas released directly from the nucleus.

\begin{table}[ht!]
  \caption{JWST/MIRI volatile measurements relevant to the panspermia interpretation. The ratios make 3I/ATLAS a useful biosignature test case, but not evidence for biology by themselves.}
  \label{tab:volatile_chronology}
  \centering
  \setlength{\tabcolsep}{3pt}
  \begin{tabular}{P{0.18\columnwidth} P{0.21\columnwidth} P{0.21\columnwidth} P{0.31\columnwidth}}
    \toprule
    Quantity & Dec. 15--16 & Dec. 27 & Why it matters \\
    \midrule
    $Q_{\rm CH4}$ & $4.2\times10^{26}$ s$^{-1}$ & $2.3\times10^{26}$ s$^{-1}$ & Methane biomass test \\
    $Q_{\rm CO2}/Q_{\rm H2O}$ & $2.30\pm0.03$ & $5.16\pm0.13$ & CO$_2$-rich relative to typical comets \\
    $Q_{\rm CH4}/Q_{\rm H2O}$ & $11.0\pm0.5\%$ & $21.6\pm1.3\%$ & CH$_4$ enriched relative to water \\
    \bottomrule
  \end{tabular}
\end{table}

\section{Natural panspermia}

\subsection{Survival in protective ice films and shielding}

Microbes can survive freezing under some terrestrial conditions, but survival is not the same as growth. The relevant natural-panspermia picture is a protected cell or biomolecule embedded in ice, dust, or a shielded pore, not an exposed organism on the surface. Rohde and Price modeled diffusion-controlled metabolism for isolated microorganisms trapped in ice crystals, where nanometer-scale liquid films and diffusion of small molecules can support repair-level metabolism rather than reproduction \citep{rohdeprice2007}. Related work on Greenland ice connected excess methane to microbial activity at low metabolic rates \citep{tung2005}. Laboratory experiments show that bacteria can incorporate DNA and protein precursors at $-15^\circ$C \citep{christner2002}, and \emph{Psychrobacter arcticus} can repair radiation-induced DNA double-strand breaks at $-15^\circ$C without net growth \citep{dieser2013}.

These results support the possibility that icy material can preserve dormant or slowly repairing cells. They do not imply that a large active biosphere exists inside every icy body. Long-term survival and transfer require shielding from radiation, a tolerable chemical environment, and enough trace energy and reactants to repair molecular damage, as emphasized in lithopanspermia studies of viable microbial transfer \citep{mileikowsky2000}. Revival or growth requires more: liquid water, chemical energy, nutrients, and enough time at a temperature compatible with metabolism.

\subsection{Does sunlight revive frozen microbes?}

If microbes are frozen inside an iceberg-like ISO and later exposed to sunlight, heating alone is not sufficient for revival. A dormant cell is not revived simply by reaching a warmer temperature; it must have liquid water and chemical resources. Solar heating near perihelion can warm the surface and shallow subsurface, drive sublimation, liberate icy grains, and perhaps create transient thin films or brines in favorable local microenvironments. It does not automatically create a sustained liquid-water interior.

The 10 K temperature floor used for dormant interstellar cruise does not apply to the sunlit surface of 3I/ATLAS near perihelion. At the perihelion distance $r\simeq1.36$ AU, a simple blackbody equilibrium temperature is
\begin{equation}
T_{\rm BB}\simeq 278~\mathrm{K}\,r^{-1/2}\simeq 239~\mathrm{K},
\end{equation}
before corrections for albedo, rotation, thermal inertia, sublimation cooling, and local geometry. Subsolar surface temperatures can be higher, while sublimation can buffer icy surfaces. The key point is that near the Sun the active surface is controlled by solar heating and volatile loss, not by the 10 K interstellar boundary.

Standard comet models generally find that solar heating strongly affects only shallow layers, often centimeters to meters, while the deeper nucleus remains insulated over a single perihelion passage \citep{prialnik2004}. Significant interior liquid water or mobile near-surface phases require additional conditions such as sufficient size, low thermal conductivity, salts, permeability, gas diffusion, or porous material properties \citep{fellows2020}. Earlier heating by short-lived radionuclides is a separate formation-history mechanism, especially heating by $^{26}$Al in cometary bodies \citep{prialnikbarnun1987}. Therefore, the conservative natural-panspermia scenario is not a warm living ocean inside 3I/ATLAS today; it is the release of preserved cells, fragments, or biomolecules from shielded ice during solar-driven activity.

\subsection{Could methanogens produce the methane?}

Methane can be biological on Earth. Strictly speaking, the strongest methane producers are not bacteria but methanogenic archaea. They can produce CH$_4$ through pathways such as CO$_2+4$H$_2\rightarrow$CH$_4+2$H$_2$O, acetoclastic methanogenesis, or methylotrophic methanogenesis. The JWST/MIRI methane rates used below were measured after perihelion, on 2025 December 15--16 and December 27, when 3I/ATLAS was already outbound rather than at closest approach. The relevant question is quantitative: how much biological material would be needed under frozen-survival conditions versus active, substrate-rich methanogenesis?

Using the PNAS survival-metabolism estimate of $1900$ CH$_4$ molecules cell$^{-1}$ yr$^{-1}$ at $-10^\circ$C, the required number of cells is
\begin{equation}
N_{\rm cell} = \frac{Q_{\rm CH4}(1~\mathrm{yr})}{1900}.
\end{equation}
This estimate is linearly sensitive to the assumed temperature-dependent per-cell rate. If the actual survival metabolism rate is $f_T$ times the PNAS value, then $N_{\rm cell}$ and the inferred biomass are both reduced by $f_T^{-1}$. For example, a $10^3$ times faster rate would lower the required biomass by $10^3$, while a $10^3$ times slower rate would raise it by $10^3$. The active-methanogenesis cases below therefore bracket the opposite, warm-liquid limit rather than a small correction to the frozen $-10^\circ$C survival rate.
For the JWST/MIRI methane rates this implies $N_{\rm cell}\sim 10^{30}$--$10^{31}$. With a typical cell mass of 38 fg, the corresponding biomass is $\sim 10^{14}$--$10^{15}$ kg (Table~\ref{tab:methane_budget}). This is the conservative frozen-survival limit, not the maximum possible biological methane rate.

The opposite limit is active methanogenesis in a warm, liquid, chemically supplied environment. Laboratory and bioreactor studies report much higher rates. For example, \emph{Methanosarcina acetivorans} growing on methanol has a measured CH$_4$ secretion rate of $22$ mmol gDW$^{-1}$ h$^{-1}$ \citep{benedict2012}; \emph{Methanocaldococcus jannaschii} under high-H$_2$ conditions reaches $496\pm21$ fmol cell$^{-1}$ h$^{-1}$ \citep{topcuoglu2019}; and \emph{Methanobacterium congolense} provides a mesophilic hydrogenotrophic benchmark of order tens of fmol cell$^{-1}$ d$^{-1}$ under CO$_2$-limited conditions \citep{chen2019}. These rates show that biological methane production is conceivable in principle if an ISO contains warm liquid microenvironments, abundant H$_2$/CO$_2$ or methylated substrates, nutrients, and enough active biomass. They do not follow from dormant ice survival alone.

Even if the free energy of methanogenesis is taken as $\sim 134$ kJ mol$^{-1}$ CH$_4$, the observed methane rate corresponds to an energy scale of only $\sim 10^8$ W; the difficulty is supplying a habitat with enough liquid water, redox reactants, nutrients, permeability, and active biomass. The JWST/MIRI analysis favors an abiotic explanation in which CH$_4$ was depleted from outer layers and then emerged from buried, less processed subsurface ice as the post-perihelion thermal wave reached deeper material \citep{belyakov2026jwst}. The biological interpretation is therefore not ruled out by reaction kinetics alone, but it requires an active habitat rather than only preserved frozen cells.

\begin{table}[ht!]
  \caption{Natural-panspermia methane budget under frozen-survival and optimistic active-methanogenesis limits. Per-cell biomass estimates use 38 fg per cell only as an order-of-magnitude scale; gDW denotes grams dry weight.}
  \label{tab:methane_budget}
  \centering
  \setlength{\tabcolsep}{3pt}
  \begin{tabular}{P{0.25\columnwidth} P{0.29\columnwidth} P{0.38\columnwidth}}
    \toprule
    Case & Rate assumption & Required cells and biomass \\
    \midrule
    Frozen ice survival & 1900 cell$^{-1}$ yr$^{-1}$ & Dec. 15--16: $7.0\times10^{30}$ cells; $2.7\times10^{14}$ kg. Dec. 27: $3.8\times10^{30}$ cells; $1.5\times10^{14}$ kg. \\
    \emph{M. congolense} active H$_2$/CO$_2$ & 48.8 fmol cell$^{-1}$ d$^{-1}$ & Dec. 15--16: $1.2\times10^{21}$ cells; $4.7\times10^{7}$ kg. Dec. 27: $6.8\times10^{20}$ cells; $2.6\times10^{7}$ kg. \\
    \emph{M. jannaschii} high-H$_2$ & 496 fmol cell$^{-1}$ h$^{-1}$ & Dec. 15--16: $5.1\times10^{18}$ cells; $1.9\times10^{5}$ kg. Dec. 27: $2.8\times10^{18}$ cells; $1.1\times10^{5}$ kg. \\
    \emph{M. acetivorans} on methanol & 22 mmol gDW$^{-1}$ h$^{-1}$ & Dec. 15--16: $1.1\times10^{5}$ kg dry wt. Dec. 27: $6.2\times10^{4}$ kg dry wt. \\
    \bottomrule
  \end{tabular}
\end{table}

Thus, microbes surviving in ice are plausible in principle, and active methanogens could in principle produce far more methane than frozen repair metabolism. The natural-panspermia result possibility implies that 3I/ATLAS-like objects could preserve dormant biology or biomolecules in shielded ice, and perihelion or outbound activity could release that material into grains or gas. A biological contribution to the observed methane becomes conceivable only if a much larger warm active habitat, a high metabolic rate, and adequate reactants existed; repair-level survival metabolism alone remains insufficient.

\section{Feasibility of Directed Panspermia}

Directed panspermia asks a different question. Instead of asking whether 3I/ATLAS naturally carried life, we ask whether a technological project could deliberately plant life-bearing material inside or onto an icy ISO like 3I/ATLAS when it passes near a star, using the ISO as a long-lived interstellar carrier. The directed scenario is not simply delivery to one ISO; it is a deliberate attempt to use the ISO population as carriers that could seed planetary systems throughout the Milky Way.

\subsection{Emplacement cannot be a direct 60 km s$^{-1}$ impact}

The most important engineering constraint is impact speed. A direct collision at $60~\mathrm{km~s^{-1}}$ releases a kinetic energy per unit mass of,
\begin{equation}
\frac{E}{m}=\frac{v^2}{2}\simeq 1.8\times10^9~\mathrm{J~kg^{-1}},
\end{equation}
which is more than 400 times the specific energy of TNT, $4.2\times10^6$ J kg$^{-1}$. At this speed, the collision produces shocks, melting, vaporization, and plasma. NASA hypervelocity-impact guidance treats meteoroid impacts up to $\sim 72~\mathrm{km~s^{-1}}$ as a structural-damage regime requiring specialized analysis rather than ordinary penetration mechanics \citep{elfer1996}. A biological capsule should not be expected to survive direct impact with the nucleus.

Therefore, a directed-panspermia architecture must avoid hypervelocity impact of the biological payload. Possible concepts include: matching velocity well enough for gentle surface deposition, releasing a penetrator only after substantial relative-speed reduction, depositing material onto porous ice or dust that later becomes buried, using a sacrificial impactor only to expose fresh ice while a separate payload arrives later, or planting material during a lower-relative-speed encounter with a bound or temporarily captured object. A DART-style impactor remains useful only if the impactor is sacrificial and the life-bearing payload is dynamically separated from the impact \citep{cheng2023}. A better biological-delivery architecture would be closer to a low-relative-speed landing or touch-and-go operation, analogous in spirit to OSIRIS-REx sampling of asteroid Bennu, rather than to a destructive kinetic impact \citep{bierhaus2018}.

The scientific goal of this scenario is not to keep the entire ISO warm. The goal is to use the ISO as radiation shielding, thermal buffering, and interstellar transport after the object leaves the planetary system. The payload could contain dormant spores, cryopreserved cells, or prebiotic/biological material embedded in an engineered matrix; Bacillus endospores are a standard reference case for resistance to extreme terrestrial and extraterrestrial stresses \citep{nicholson2000}. Success requires three separate steps: delivery to the ISO without sterilizing shock, emplacement into protected ice or porous material, and return to a cold long-term cruise state after the active perihelion phase.

\subsection{Thermal requirements for a biological payload}

A directed payload must preserve viability before and after emplacement. For a spherical capsule of radius $R$, projected area $A_p=\pi R^2$, and emitting area $A_e=4\pi R^2$, the heater power needed to hold temperature $T$ is
\begin{equation}
P_{\rm heat}=\max\left[0,A_e\epsilon\sigma T^4-A_p\alpha\frac{S_0}{r^2}\right],
\end{equation}
where $\epsilon$ is emissivity, $\alpha$ is absorptivity, $S_0$ is the solar constant, and $r$ is heliocentric distance in AU. For a 0.2 m capsule held near 283 K, the earlier engineering estimate gives $\sim 11$ W at 1 AU and $\sim 150$--165 W beyond 3 AU for a bare surface, or $\sim 1$ W at 1 AU and $\sim 9$--10 W beyond 3 AU for a low-emissivity/low-absorptivity insulated case.

This heater budget applies to the spacecraft or payload, not to the ISO nucleus. A payload may be kept warm by onboard power, radioisotope heater units, insulation, or mission timing; NASA radioisotope heater units provide about one watt of heat each \citep{nasaRHU}. This does not imply that the ISO interior is habitable.

\begin{table}[ht!]
  \caption{Directed-panspermia feasibility gates for an ISO like 3I/ATLAS.}
  \label{tab:directed_gates}
  \centering
  \fitwidth{%
  \begin{tabular}{l l}
    \toprule
    Requirement & Implication \\
    \midrule
    Gentle emplacement & Direct 60 km s$^{-1}$ impact destroys payload \\
    Thermal protection & Payload heating/insulation is separate from ISO heating \\
    Shielding & Ice or porous material must reduce radiation damage \\
    Liquid water after delivery & Needed for revival, not for dormant transport \\
    Long-term cruise & Payload should return to a cold, low-metabolism state \\
    \bottomrule
  \end{tabular}}
\end{table}

\section{Shared Thermal Constraints for Natural and Directed Panspermia}

\subsection{Dormant interstellar cruise}

For a dormant ISO moving through the interstellar medium, we treat the surface temperature as an environmental boundary condition. \citet{hoangloeb2020} discuss heating and destruction processes for icy interstellar bodies, including interstellar radiation, gas and dust impacts, cosmic rays, and collisional heating. We use that work only as support for the relevant heating channel; the specific dormant-cruise scaling below is the order-of-magnitude balance between gas-collisional energy input and thermal emission,
\begin{equation}
T_{\rm HL}\approx 273~\mathrm{K}
\left(\frac{n_{\rm H}}{10~\mathrm{cm^{-3}}}\right)^{1/4}
\left(\frac{v}{0.1c}\right)^{3/4}.
\end{equation}
This follows from equating the incident gas-heating flux, proportional to $n_{\rm H}m_{\rm H}v^3$, to radiative cooling, proportional to $\sigma T^4$.
The velocity normalization is not the adopted ISO speed. For a typical ISO speed $v=30~\mathrm{km~s^{-1}}\simeq 10^{-4}c$ and $n_{\rm H}=1~\mathrm{cm^{-3}}$, this gives $T_{\rm HL}\approx 0.9$ K. Even at $v=60~\mathrm{km~s^{-1}}$, the estimate remains only $\approx 1.5$ K. In this paper, the dormant-cruise surface boundary is therefore represented as
\begin{equation}
T_s=\max(T_{\rm HL},T_{\rm CMB},T_{\rm floor}),\quad T_{\rm floor}=10~\mathrm{K},
\end{equation}
where the floor is a conservative bundle for unresolved environmental heating. This is not a perihelion temperature.

Present-day radiogenic heating then enters only as an interior source. For uniform volumetric heating $q_{\rm vol}=\rho q_{\rm rad}$ and conductivity $k$,
\begin{equation}
T(r)=T_s+\frac{\rho q_{\rm rad}}{6k}(R^2-r^2).
\end{equation}
For $R=100$ m, present-day radiogenics raise the center by only $\sim 2\times10^{-3}$ K for granite-like rock and $\sim 10^{-6}$ K for an icy mix. A $\sim 300$ K center with a 10 K surface requires a radius of order 40 km for granite-like present-day radiogenic heating, and much larger for icy material.

\begin{table}[ht!]
  \caption{Radiogenic specific power used to test whether an ISO can maintain a warm interior during dormant cruise. These are interior source terms only; they do not set the perihelion surface temperature.}
  \label{tab:radiogenic_power}
  \centering
  \fitwidth{%
  \begin{tabular}{l c}
    \toprule
    Material & Specific power (W kg$^{-1}$) \\
    \midrule
    Granite-like rock & $1.1\times10^{-9}$ \\
    Igneous range & $4.2\times10^{-11}$--$3.6\times10^{-9}$ \\
    Pure H$_2$O ice, tritium-limited & $1.1\times10^{-16}$ \\
    Pure CO$_2$ ice, $^{14}$C upper bound & $1.6\times10^{-12}$ \\
    50/50 H$_2$O/CO$_2$ mix & $8.0\times10^{-13}$ \\
    \midrule
    Granite / water ice & $\sim 10^7$ \\
    Granite / 50/50 ice mix & $\sim 1.4\times10^3$ \\
    \bottomrule
  \end{tabular}}
\end{table}

\begin{table}[ht!]
  \caption{Center--surface temperature excess from present-day radiogenics with $T_s=10$ K. Rocky: $\rho=2650$ kg m$^{-3}$, $k=2.5$ W m$^{-1}$ K$^{-1}$, $q_{\rm rad}=1.1\times10^{-9}$ W kg$^{-1}$. Icy mix: $\rho=900$ kg m$^{-3}$, $k=0.5$ W m$^{-1}$ K$^{-1}$, $q_{\rm rad}=8.0\times10^{-13}$ W kg$^{-1}$.}
  \label{tab:thermal_sweep}
  \centering
  \fitwidth{%
  \begin{tabular}{r r c c}
    \toprule
    Radius $R$ (m) & $R$ (km) & $\Delta T_{\rm rock}$ (K) & $\Delta T_{\rm icy}$ (K) \\
    \midrule
    50 & 0.05 & $4.9\times10^{-4}$ & $6.0\times10^{-7}$ \\
    100 & 0.10 & $1.9\times10^{-3}$ & $2.4\times10^{-6}$ \\
    1000 & 1 & $1.9\times10^{-1}$ & $2.4\times10^{-4}$ \\
    $10^4$ & 10 & $1.9\times10^{1}$ & $2.4\times10^{-2}$ \\
    $10^5$ & 100 & $1.9\times10^{3}$ & $2.4$ \\
    \bottomrule
  \end{tabular}}
\end{table}

\subsection{Active perihelion}

Near perihelion, the dormant-cruise model is not the relevant surface model. Solar heating can drive H$_2$O, CO$_2$, CO, and organic release; entrained icy grains can carry volatile-rich material into the coma; and newly exposed subsurface reservoirs can explain delayed species such as methane. The panspermia question near perihelion is therefore not whether the entire nucleus becomes habitable, but whether activity exposes preserved biological material or creates short-lived microenvironments with liquid films or brines.

\section{Results organized by panspermia scenario}

\subsection{Natural panspermia results}

The natural-panspermia case has three constraints. First, microbial survival in ice is physically plausible at repair-level metabolism, especially in liquid films, veins, or frozen matrices. Second, revival or growth requires liquid water, nutrients, and energy; solar heating near perihelion may help at shallow depths but does not automatically create a sustained liquid-water interior. Third, the post-perihelion methane budget depends strongly on the assumed biological state. Frozen survival-level methanogenesis would require $\sim 10^{30}$--$10^{31}$ cells and $\sim 10^{14}$--$10^{15}$ kg of biomass, but active methanogenic archaea in warm, substrate-rich liquid environments can produce methane many orders of magnitude faster. Therefore, natural panspermia remains plausible as preservation and release of dormant material, and a biological contribution to methane is conceivable in principle, but only if 3I/ATLAS had active liquid-water niches with sufficient reactants and biomass.

\subsection{Directed panspermia results}

The directed-panspermia case is limited by dynamics and payload survival. A direct 60 km s$^{-1}$ capsule impact would destroy or sterilize the payload. Directed seeding must instead use gentle emplacement, a separated payload, or another low-relative-speed architecture. Thermal control of a payload is plausible at the watt-to-hundreds-of-watts scale depending on insulation and heliocentric distance, but warming the payload is not the same as warming the ISO nucleus. The mission problem is therefore to protect and emplace life-bearing material inside shielding ice or porous material, then let the ISO carry that protected payload through interstellar space.

\subsubsection{Mission opportunity constraints}

Survey and reachability calculations remain relevant only because a directed or diagnostic mission requires a target. Here $q$ denotes the differential size-distribution index, $dN/dD\propto D^{-q}$, where $D$ is object diameter. The current $q=2.5$ pipeline run should be read as a shallow-slope sensitivity case, not as a measured ISO size distribution. \citet{bottke2005} are used only as an asteroid-population analogy showing that relatively flat power-law slopes occur in some small-body diameter ranges. With this assumption, a Rubin/LSST-style 10 yr survey gives an order-of-magnitude baseline of $\sim 382$ detections above 100 m diameter after rescaling; the assumed wide-fast-deep survey scale is motivated by the LSST reference design \citep{ivezic2019}. Median 50\% completeness distances are 3.0, 15.0, 30.3, and 136.6 AU for 100 m, 500 m, 1 km, and 5 km diameter thresholds. For a 0.1 AU cross-path miss distance, the same run gives reachable fractions of $\sim 7$--20\% for 2--4 km s$^{-1}$ lateral maneuvers and $\gtrsim 95\%$ for 12--15 km s$^{-1}$. These numbers support mission opportunity estimates, but they are secondary to the biological and thermal feasibility constraints.

\begin{table}[ht!]
  \caption{Mission-opportunity calculations that support, but do not drive, the panspermia argument. Detection distances and counts use the $q=2.5$ Rubin-style 10 yr survey baseline as a shallow-slope sensitivity case. Here $d_{50}$ is the heliocentric distance at which the detection completeness is 50\%. Reachability uses a 0.1 AU cross-path miss distance.}
  \label{tab:mission_opportunity}
  \centering
  \setlength{\tabcolsep}{3pt}
  \begin{tabular}{P{0.39\columnwidth} P{0.18\columnwidth} P{0.33\columnwidth}}
    \toprule
    Quantity & Value & Use in this paper \\
    \midrule
    $d_{50}$ for 100 m objects & 3.0 AU & Warning distance for small ISOs \\
    $d_{50}$ for 500 m objects & 15.0 AU & Warning distance for Borisov-scale objects \\
    $d_{50}$ for 1 km objects & 30.3 AU & Warning distance for ATLAS-scale objects \\
    $d_{50}$ for 5 km objects & 136.6 AU & Rare large carrier targets \\
    Detections $\geq$100 m & 382.1 per 10 yr & Target supply estimate \\
    Reachable at 2--4 km s$^{-1}$ & 7--20\% & Low maneuver feasibility \\
    Reachable at 12--15 km s$^{-1}$ & $\gtrsim95\%$ & High maneuver feasibility \\
    \bottomrule
  \end{tabular}
\end{table}

Detailed class-normalized counts are moved to Appendix~\ref{app:mission_details} so that the main text keeps the requested natural/direct panspermia structure.

\section{Limitations}

This paper evaluated prospects, not proof of life in interstellar objects. The methane analysis is an order-of-magnitude budget and does not model spatially resolved coma chemistry, isotope ratios, or all abiotic methane pathways. The microbial survival literature is terrestrial and cannot be transferred directly to 3I/ATLAS without assumptions about shielding, radiation dose, salts, nutrients, and cell type.

The thermal model is deliberately split into limiting regimes. The dormant-cruise calculation uses an environment-fixed surface and steady conduction; it does not include porosity, gas transport, amorphous ice, crystallization, brines, clathrate kinetics, or formation-era heating in detail. The perihelion discussion is not a full thermophysical model of 3I/ATLAS. A future version should calculate time-dependent solar heat penetration and possible transient liquid films or brines for the observed perihelion distance and activity history.

The directed-panspermia mission concept is also preliminary. It identifies the fatal problem with direct hypervelocity capsule impact, but it does not yet design a complete low-relative-speed emplacement architecture, contamination-control protocol, or long-term viability strategy.

\section{Conclusions}

The prospects for panspermia via 3I/ATLAS-like ISOs divide naturally into natural and directed scenarios. Natural panspermia is plausible as long-term preservation of cells or biomolecules inside shielded ice. The observed methane production is too large to be explained by frozen survival-level methanogenesis, but active methanogenic archaea could in principle produce much more methane if warm liquid habitats and suitable reactants existed. The correct observational posture is therefore to treat methane and organics as diagnostics, not as evidence for life.

Directed panspermia has different feasibility challenges. It may be possible in principle to use an icy ISO as an interstellar carrier, but the biological payload cannot be delivered by a direct 60 km s$^{-1}$ impact. Any realistic architecture must place the payload gently, analogous to the OSIRIS-REx touch-and-go sampling of asteroid Bennu, preserve it thermally, and exploit the ISO as shielding and transport rather than relying on the ISO to become habitable \citep{bierhaus2018}.

The strongest conclusion is that 3I/ATLAS-like objects are valuable because they make panspermia testable. Observations of volatile timing, organics, methane, isotopes, chirality, and dust/ice grains can constrain natural panspermia. Mission studies of gentle emplacement and payload survival can constrain directed panspermia. Both questions require separating dormant interstellar cruise, active perihelion, and engineered payload physics.

\appendix

\section{Supplementary Mission-Opportunity Details}
\label{app:mission_details}

We also compute object-class normalizations for Oumuamua-like, Borisov-like, and ATLAS-like populations. The calculation uses the empirical class densities and size bands as anchors: Oumuamua-like objects use $n=0.2~\mathrm{AU^{-3}}$ in the 100--200 m diameter band with $v_{\rm rel}=26~\mathrm{km~s^{-1}}$; Borisov-like objects use $n=0.03~\mathrm{AU^{-3}}$ in the 500--1000 m band with $v_{\rm rel}=44~\mathrm{km~s^{-1}}$; and ATLAS-like objects use $n=0.01~\mathrm{AU^{-3}}$ in the 1000--5000 m band with $v_{\rm rel}=60~\mathrm{km~s^{-1}}$. For each class we infer the total density implied by the fraction of the $q=2.5$ size distribution in that band, cap the total density at $0.1~\mathrm{AU^{-3}}$, and then compute cumulative 10 yr detection counts below.

\noindent\textit{Class-normalized detection counts.}
\begin{center}
  \fitwidth{%
  \begin{tabular}{l c c c c}
    \toprule
    Class & $\geq$100 m & $\geq$500 m & $\geq$1 km & $\geq$5 km \\
    \midrule
    Oumuamua-like & 1155.8 & 1056.3 & 980.2 & 652.8 \\
    Borisov-like  & 1956.0 & 1787.7 & 1658.8 & 1104.7 \\
    ATLAS-like    & 2667.3 & 2437.7 & 2262.0 & 1506.4 \\
    \bottomrule
  \end{tabular}}
\end{center}

\begin{center}
  \includegraphics[width=\hsize]{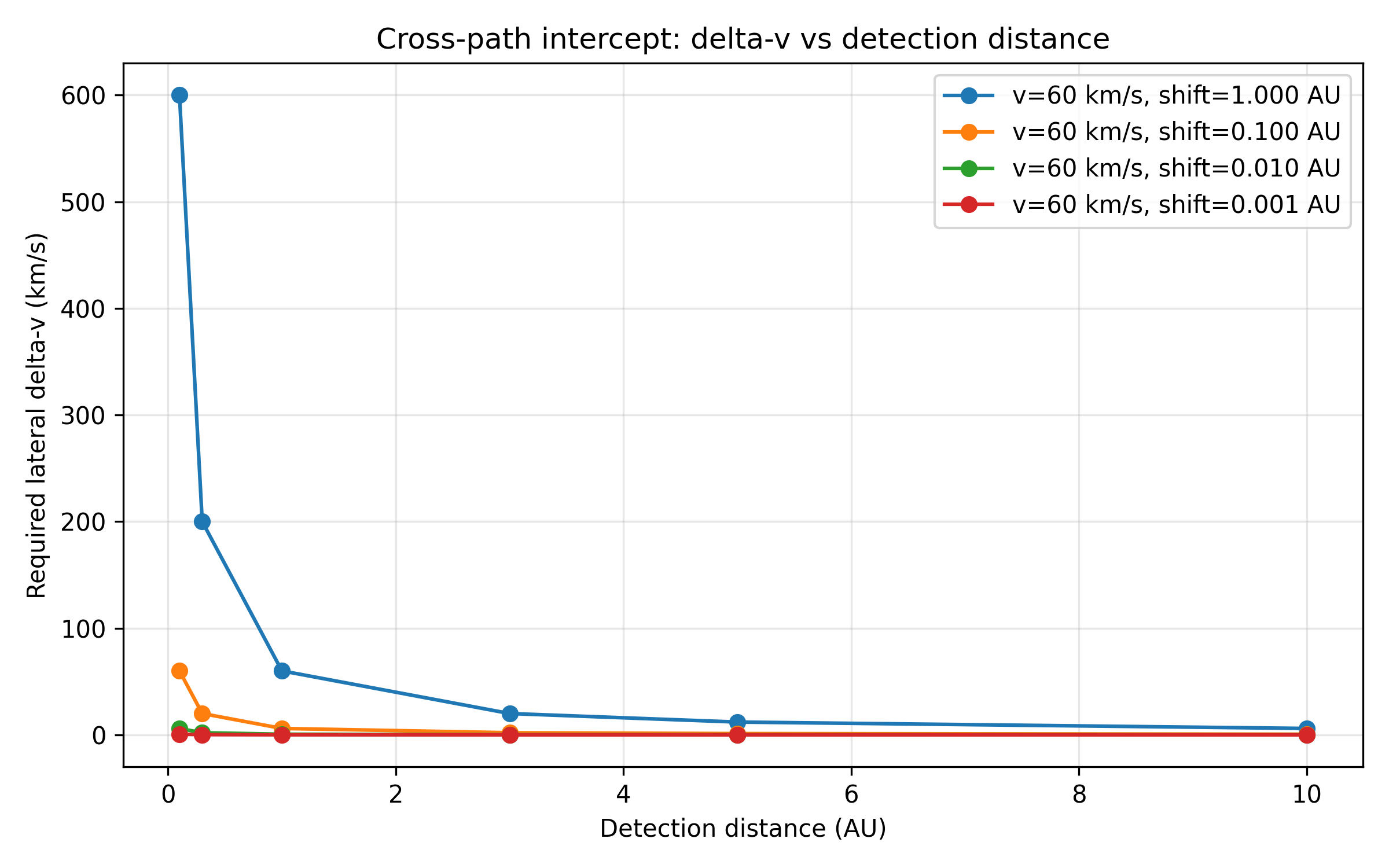}
\end{center}
\noindent\textit{Cross-path maneuver requirement versus detection distance. This plot is retained only as a directed-panspermia mission-opportunity constraint: it estimates whether a spacecraft can reach the ISO path. It does not imply that a biological payload can survive direct impact with the nucleus.}

\end{document}